# Emergence of new optical resonances in single-layer transition metal dichalcogenides with atomic-size phase patterns

*John M. Woods[1][†], Saroj B. Chand[1][†], Enrique Mejia[1], Takashi Taniguchi[2], Kenji Watanabe[3], Johannes Flick[4,5,6], Gabriele Grosso[1,6]\**

1. Photonics Initiative, Advanced Science Research Center, City University of New York, New York, NY, USA

2. International Center for Materials Nanoarchitectonics, National Institute for Materials Science, 1-1 Namiki, Tsukuba 305-0044, Japan

3. Research Center for Functional Materials, National Institute for Materials Science, 1-1 Namiki, Tsukuba 305-0044, Japan

4. Center for Computational Quantum Physics, Flatiron Institute, New York, NY 10010, USA

5. Department of Physics, City College of New York, New York, NY 10031, USA

6. Physics Program, Graduate Center, City University of New York, New York 10016, New York, USA

† Equal contribution

E-mail: ggrosso@gc.cuny.edu

**Atomic-scale control of light-matter interactions represent the ultimate frontier for many applications in photonics and quantum technology. Two-dimensional semiconductors, including transition metal dichalcogenides, are a promising platform to achieve such control due to the combination of an atomically thin geometry and convenient photophysical properties. Here, we demonstrate that a variety of durable polymorphic structures can be combined to generate additional optical resonances beyond the standard excitons. We theoretically predict and experimentally show that atomic-sized patches of 1T phase within the 1H matrix form unique electronic bands that give rise to new and robust optical resonances with strong absorption, circularly polarized emission and long radiative lifetime. The atomic manipulation of two-dimensional semiconductors opens unexplored scenarios for light harvesting devices and exciton-based photonics.**

Light–matter interactions are the underlying mechanisms of a wide range of scientific disciplines and they have unlocked a series of key technological advancements. Controlling such interactions at the atomic scale enables the exploration of new physics and allows unprecedented levels of device scalability. However, engineering atomic-scale systems is challenging and requires suitable material platforms. Transition metal dichalcogenides (TMDs) have proven to be an ideal platform for a variety of applications in photonics, electronics, and chemistry due to their malleability resulting from the atomically-thin geometry combined with convenient electronic band structures.(*1*, *2*) The latter leads to robust excitonic resonances imbued with spin-valley locking, large oscillator strength, and large binding energy. In the semiconducting (1H) phase, monolayers of TMDs have a bandgap in the visible range of the spectrum that give rise to several bright(*3*, *4*) and dark exciton

states.(*5–7*) The splitting of the valence band resulting from the large spin-orbit coupling (SOC) generates two exciton resonances at the *K* symmetry point of the Brillouin zone. These are the well-known *A* and *B* excitons. The energy difference between these two excitons depends on the type of TMD, with the *A* excitons being the excitonic ground state of the system. Another peak, corresponding to the *C* exciton, appears in the absorption spectrum of TMDs and emerges from a band nesting effect.(*8–10*)

Besides optical properties, TMDs have attracted growing interest for their mechanical and chemical properties, including polymorphism. For example, transition-metal disulfides, such as $MoS_2$ and $WS_2$, primarily exist in the thermodynamically favored semiconducting 1H phase, but can be forced to the 1T phase with metallic properties through bottom-up growth(*11, 12*) or top-down treatment(*13, 14*) pathways. In the 1H phase, both S atoms are located on top of each other (when viewed along the c-axis) resulting in a trigonal prismatic symmetry about the metal atom, whereas in the 1T phase the two S atoms are displaced into an octahedral symmetry, as shown in Figure S1. Transitions from the 1H to the 1T phase (1H→1T) can be triggered by weak external perturbations, such as chemical, thermal, and irradiation treatments. Phase boundaries have found several applications in electronics and catalysis.(*15–17*) However, the optical properties and the possible photonic applications of phase boundaries, mixed-phase, and atomic-sized phase structures in two-dimensional materials have not been exhaustively explored yet.

Here, we demonstrate that new optical transitions can occur across the 1H/1T phase boundaries, in which the metallic 1T phase provides electrons that are optically excited towards an unoccupied conduction band of the semiconducting 1H phase. By using monolayers $WS_2$ as blank canvas, we sculpt mixed-phased structures with atomic-size patterns and show that additional optical resonances appear in parts of the spectrum that are naturally inaccessible in pristine 1H materials. We unveil the emergence of a new exciton peak in both absorption and emission spectra, which we refer to as the *M* exciton because it stems from the mixed phase structure. Ab-initio calculations based on density functional theory (DFT) pinpoint the orbital origin of the *M* exciton and micro-photoluminescence experiments reveal promising optical properties, including room temperature stability, large oscillator strength, long lifetime, and circular polarization. The demonstration of the creation of new resonances through atomic manipulations of TMDs paves the way towards the engineering of their full optical spectrum and the realization of a new generation of atomic-size devices.

**Atomic-size phase patterns in WS$_2$**

In TMDs, the 1H →1T phase transition occurs through the collective displacements of atoms and involves stable intermediate structures that eventually merge together forming the final 1T atomic lattice geometry.(*18*) When driven by external stimuli, this phase transition can be interrupted midway generating mixed-phase states in which atomic-size 1T grains are assembled in the 1H matrix with peculiar phase boundaries. The size of the grains can be controlled by tuning the duration and the strength of the external stimulus used to trigger the phase transition. We engineer phase mixture states by irradiating monolayers of WS$_2$ with argon plasma. This technique has previously been used to achieve a full metallic phase and to macroscopically pattern different phases in TMDs.(*13*) In our experiments, the plasma power and the irradiation time are tuned such that the treatment does not fully convert the pristine 1H monolayers (see Methods), but generate instead 1T structures of different shapes and sizes embedded within the broader 1H phase, as shown in Figure 1. Figure 1a-c illustrates the atomic models of the intermediate steps of the phase transition achieved by plasma irradiation, while Fig. 1d-f are simulations of how the same structures would look in atomically resolved electron images. Figures 1g-i show experimental images of the atomic lattice of a pristine monolayer (Fig. 1g) and plasma treated monolayers with an irradiation time of 5 s (Fig. 1h) and 10 s (Fig. 1i). Atomically resolved images of the lattice are obtained by high-resolution transmission electron microscopy (TEM) taken with focusing conditions for bright atom contrast that allow us to clearly resolve W and S atoms, as well as S vacancies and different phases. Line profiles (taken along the direction indicated by yellow arrows in Fig.1d-i) allow us to identify the 1H and 1T phase regions.(*14*, *19*, *20*)

The effect of the plasma is two-fold: it creates S vacancies and, at the same time, provides the shear force and energy responsible for the shift of the S atoms out of their thermodynamically favored 1H positions towards the 1T phase.(*14*, *18*, *21*) Analysis of experimental TEM images of sample irradiated for 5 s reveals the presence of intermediated, linear 1T structures with lateral dimension of 1 lattice vector near a line of S vacancies. Similar structures were identified as precursors of the 1T phase in the pioneer studies on electron-driven phase transitions in TMDs.(*18*) Simulated TEM images of a linear 1T structure within 1H-$WS_2$ (Fig. 1e) are consistent in appearance and show a similar line-profile to the experimentally observed 1T grain in Figure 1h. To differentiate from other structures we will refer to these 1T grains with linear morphologies and varying length as 1T-L1. A detailed discussion on the characterization of these structures is in Figure S2. Less frequently, we observe in the samples treated for 5s the presence of small 1T triangular grains. Longer treatments give rise to larger triangular or polygonal grains of 1T phase surrounded by the 1H phase (Fig.1i). We refer to the triangular grains as 1T-Tr*N*, where *N* indicates the lateral size in lattice vectors of the 1T triangle. TEM images of plasma treated samples taken at lower magnification showing further examples of 1T-L1 and 1T-Tr structures can be found in Figure S3. Moreover, electron diffraction patterns (Figure S4) taken from a large area (~18,000 $nm^2$) of the plasma treated sample maintain the 6-fold symmetry of pristine 1H-$WS_2$ indicating both the absence of high-angle grain boundaries due to a perfect register of the induced 1T phases to the remaining regions of 1H phase, as well as further confirming the new phase is the 1T phase, and not the 1T' phase (extensively studied in Telluride TMDs), which would instead have a rectangular diffraction pattern.(*22*) The partial phase transition with plasma irradiation and copresence of 1H and 1T phases in the treated samples are further confirmed by Raman and X-Ray photoelectron (XPS) spectroscopy (Supplementary Note 1).

**Band structure and optical properties of mixed-phase patterns**

The atomic-size lateral confinement of the 1T phase within the 1H areas dramatically alters the electronic and optical properties of $WS_2$ due to the formation of new electronic states within the standard optical band gap of the 1H phase. The results of ab-initio calculations using density-functional theory on 9x9x1 supercells containing 1T/1H mixed phase structures similar to the ones observed in the TEM images are shown in Figure 2. First, we look at the band structure (obtained using band-folding, see methods) of the 1T-L1 (Fig.2a-b) and 1T-Tr2 (Fig.2c-d) within a 1H matrix. The presence of 1T-L1 or 1T-Tr2 within the 1H phase introduces new occupied electronic states in $WS_2$ just above the 1H valence band edge. These states together with the 1H K-valley conduction band host a new optical transition with large dipole moment at an energy below that of the *A* exciton. The corresponding orbitals (or Kohn-Sham wave functions) in real space for this pair of states in the 1T-Tr2 supercell is shown in Figure 2e-f. Clearly, the occupied state of the transition (Fig.2f) is localized around the triangular region and is therefore primarily attributed to the 1T phase. The electrons provided by the metallic phase can be photoexcited towards unoccupied states (Fig.2e) whose orbitals are delocalized in the 1H semiconducting phase. The shape and distribution of this orbital, in addition to its location in the electronic band structure, allow us to identify this state as the conduction band edge at the K valley of the 1H phase (Figure S8). Calculations performed on 1T-L1 return similar results regarding the nature of the orbitals (Figure S9).

Figure 2g compares the calculated imaginary part of the dielectric function for a full 1H supercell, the 1T-Tr2, and the 1T-L1 structures indexed to the energy of the *A* exciton transition. The higher energy peak is the *B* exciton, emerging from the splitting of the band edges at the K-valley due to SOC. In the 1T-Tr2 and 1T-L1 structures (red and maroon curves, respectively), another resonance, with strength comparable to the *A* and *B* excitons ($X_A$ and $X_B$), appears at around 200 meV below the *A* exciton. This new resonance is related to the occupied midgap states emerging from the 1H/1T mixed phase structure (Fig. 2f) as previously discussed, and we refer to it as the *M* exciton ($X_M$). For the sake of stability and for a realistic comparison with the experimental observations and the known role of vacancy formation to the 1H→1T phase transformation,(*14*) we use an atomic 1T-L1 structure that contains a line of vacancies at one of the two boundaries between the 1H and 1T phase. The relaxation of 1T-L1 and 1T-Tr2 structures without vacancies resulted in a return to a fully pristine 1H supercell. Remarkably, we note that these new bands that give rise to the *M* exciton are unique of the mixed-phase structures and cannot be ascribed to the constituent 1H and 1T phases, or the lines of vacancies, necessary to form a stable structure (Figure S10). In the calculated band structures, other localized, but unoccupied, states appear within the gap and they can be associated with vacancies and the small 1T grains.

We perform optical experiments on samples containing mixed-phase structures after transferring thin layers (∼20 nm) of hexagonal Boron Nitride (hBN) on top of the treated $WS_2$ monolayers. Top encapsulation results in the narrowing of the emission lines(*23*) and, at the same, suggests that the mixed-phase samples can be integrated into more complex heterostructures. Differential reflectivity spectra in Fig. 2h measured at cryogenic temperature (T=8K) for pristine and 5s-treated samples confirm the generation of an additional optical resonance. While pristine $WS_2$ shows the typical absorption spectra of the 1H phase in this energy range with the *A* exciton and *B* exciton peaks, the treated samples hosting mixed-phase atomic structures reveal the appearance of pronounced absorption peak of the *M* exciton at ~200 meV below the *A* exciton. The absorption of the *M* exciton dominates the one of the *A* exciton in treated samples. We attribute the decrease of absorption of the *A* and *B* excitons in treated samples with respect to the pristine case to the reduction of the coverage of the 1H phase as a consequence of the formation of 1T patches, as well as an increased defect density in the atomic lattice upon plasma irradiation.(*24*, *25*) The good agreement between the theoretical and experimental value of the energy of the *M* excitons relative to the *A* exciton indicates that this new peak is associated with the new mixed-phase atomic structures. The broad linewidth of the *M* exciton absorption peak can be attributed to the presence in the 5s-treated samples of multiple mixed-phase structures with different shape and interface. In fact, we observe a predominant presence of 1T-L1 structures and a minority of small triangles (Figure 1h and Figure S3). As discussed below, different structures give rise to *M* transitions with varying energy. The emission and differential reflectivity spectra of the pristine and 5s-treated samples at room temperature are shown in Figure 2i. At room temperature, the intensity of the differential reflectivity of the *A* and *M* peaks is comparable around 0.6, while the area of the absorption peak associated with the *M* excitons is larger than the *A* exciton suggesting a stronger oscillator strength for this new transition.(*26*) Furthermore, the emission intensity of the *M* exciton dominates the one of the *A* exciton in the PL spectrum, indicating that the *M* exciton represents the preferred relaxation pathways in mixed-phase samples.

Having confirmed the emergence of a new optical resonance in mixed-phase structures, a discussion on how the optical transition occurs across the two phases is in order. Although the occupied and unoccupied states giving rise to the *M* excitons show mostly 1T and 1H character, respectively, a weak hybridization of the electronic bands occurs at the 1H/1T interface. Moreover, the spatial extent of the transition density, $\psi_{VB}^{\dagger}(\vec{r}) \cdot \psi_{CB}(\vec{r})$, shown in Figure 3a indicates that the overlap is greatest at the interface where the optical transition is most likely to take place. In contrast, the transition density for the $X_A$ transition (Figure S11) is delocalized across the 1H regions of the supercell. The role of the interfaces is further confirmed by the calculation of the strength of the transition dipole moments as a function of the supercell size. Figure 3b shows that while the strength of the *A* exciton in a 1T-Tr2 mixed-phase structure increases with increasing the supercell size, the strength of the *M* exciton is almost constant.

We then characterize the optical properties of samples treated for different times, therefore hosting different atomic mixed-phase structures. Figure 3c shows the emission spectrum at T = 8K of pristine, 3s-, 5s-, and 10s-treated $WS_2$ monolayers taken at equivalent excitation power. In pristine $WS_2$, we identify the usual emission from *A* exciton complexes that appear in n-doped samples that include in descending energy the neutral exciton, triplet and singlet trions, charged biexcitons, and complexes related to dark excitons.(*5*, *6*, *27*) An extensive discussion on the attribution of these peaks is in the Supplementary Note 3. We observe that the emission intensity of all the *A* exciton complexes reduces as a function of the treatment time due to a progressive reduction of the 1H phase and the formation of the 1T grains. At 5s, emission from the *M* excitons, peaking around 165 meV below the neutral *A* exciton, becomes the dominant feature in the spectra indicating a large density of 1T-L1 and 1T-Tr2 structures in the samples. The emission spectrum of the 10s-treated sample shows the strongest emission at an even lower energy. As observed in the TEM images in Fig. 1, the mixed phase structure switches from linear and small triangular 1T grains to larger triangular 1T grains when the treatment time is increased to 10s. To understand the role of the 1T grain size, we perform a theoretical analysis on supercells containing triangle 1T grains of increasing size (see Figure S12 for the atomic structures). When we extract the energy and transition dipole strength of the *M* transitions in these structures, Figure 3d, we observe an overall redshift and strength reduction when increasing the size of the 1T triangles. Figure 3e summarizes the evolution of the *M* excitons for these larger triangle cells as compared to the 1T-L1 and 1T-Tr2 mixed phased structures. These results are in good agreement with the experiments and indicate that the energy shift seen for the samples with the longest treatment time is due to the change in the size of the 1T grains. Moreover, Fig.3e shows that each mixed-phase structure generates more than one new resonance with different energies and strengths depending on the phase interfaces involved in the atomic structure. In fact, while 1T triangular structures are separated from the 1H matrix by only one kind of phase interface, 1T lines contain two inequivalent phase interfaces on either side of the 1T grain with different arrangement of S atoms around border W atoms (see Figure S13). The presence of distinct boundaries on either side of 1T-L1 explains the observation in Fig. 3e of two *M* resonances with the orbitals illustrated in Figure S9. Except for 1T-Tr1, the phase boundaries of 1T-Tr$N$ change with the relative orientation of the triangle within the 1H matrix. The 1T-Tr2 structure in Fig. 2 has Interface-I (Fig. S13a) edges, whereas Figure S14 in the SI reports the results of the calculations for a 1T-Tr2 structure with the opposite orientation, Interface-II (Figure. S13b), with respect to the 1H phase. For this structure, the *M* exciton appears at a higher energy compared with the Interface-I 1T-Tr2 structure shown in Fig.2. In Figure 3e we differentiate the results of theoretical calculations for 1T-Tr$N$ for both orientations with a dashed line border representing the Interface-II structures. We note that the energy dependence on the orientation and size of the 1T phase relative to the 1H matrix explains the broad linewidth of the *M* resonance experimentally observed in our treated samples that contain diverse atomic-size structures.

**Optical characterization of the *M* exciton**

Although the *M* resonance peak appears in the emission spectrum with an energy comparable to the one associated with the defective band,(*28*, *29*) the former has a fundamentally different nature. Opposed to the defective band that stems from *A* excitons that, upon absorption of light, get bound to impurity or lattice defects, the new peak has a different origin as indicated by the fact that it directly absorbs light. We note that localized emission from *A* excitons bound to defects cannot be ignored in our samples due to the presence of vacancies as a consequence of the plasma treatment.(*14*, *18*, *21*) We performed power-dependent PL measurements on the treated samples to further differentiate between the *M* exciton and typical defective emission. In these measurements the intensity of an optical feature's emission can be fit to a power function: $I(P) = aP^k$, where the exponent, *k*, should follow the classification of the underlying radiative process. Exciton-like recombination is expected to exhibit a power law with $k \approx 1$, recombination from bound states involving defects with $k < 1$, and from biexciton with $k > 1$.(*30*, *31*) The power evolution of our pristine $WS_2$ sample is shown in Figure S15. In Fig.4a we report the evolution of the low-temperature emission spectrum for a 5s-treated sample as a function of the laser power over four orders of magnitude. At high power, *M* excitonic emission does not saturate and becomes the dominant emission pathway for photoexcited carriers. To quantify the emergence of the *M* exciton, we extract the PL intensity of the spectra in the band from 2.07 to 2.10 eV to track the emission of the neutral *A* exciton and from 1.80 to 1.97 eV to track the lower energy features in the sample as a function of the pump power, shown in Fig.4b with red and blue circles respectively. At low to medium-high powers where the neutral *A* exciton is apparent and not yet lost in the shoulder of trionic and biexcitonic emission, the fit returns $k_A$=0.98±0.02 for *A* excitons as expected. In the low energy range, we can distinguish two regimes. For pump powers at or below 25 μW, the fit returns $k_{Def}$=0.81±0.02 suggesting that this emission is due to defect-localized emission of the *A* exciton.(*29*) At higher pump powers, the fit returns $k_M$=1.02±0.01 indicating that the predominant low energy emission is different from the defective band. This evidence shows that this low-energy emission peak is associated with an excitonic transition. We additionally note that while the remaining 1H in the sample has defects, such as vacancies induced by the plasma treatment, the degree of defectiveness is not so extreme as the neutral exciton can still be individually resolved and is only moderately broadened by irradiation (FWHM of gaussian fit of $X_A$ increases from 11.0±0.3 to 15.5±1.1 meV with 5s irradiation time, see Figure S16). We observe a reduction of the spectral weight of the *A* excitons with respect to the one of the *M* exciton when the treatment time is increased (Figure S17). This is expected as longer plasma irradiation leads to further growth of 1T phase while decreases the remaining 1H content and also quenches overall *A* emission through the formation of additional sulfur vacancies.(*24*) The rise of the defect density as a function of irradiation time is further confirmed by the growth of the saturation coefficient for the defective band ($k_{Def}$ shown in Fig. 4c). When we treat the sample further, the *A* exciton in PL emission practically vanishes, and 1H and 1T Raman modes are very weak, indicating that the remaining 1H phase is of poor optical quality (Figure S18). However, we still see non-saturating PL emission from the *M* exciton resonance. The non-saturating photoluminescence emission ascribed to the *M* exciton is additionally observed in room temperature measurements (Figure S19), confirming the robustness of this optical transition.

The emergent *M* exciton can be additionally distinguished from defect-localized emission through fluorescence lifetime measurements as illustrated in Fig.4d. While the short lifetime of the *A* exciton is of the order of few ps(*32*) and cannot be resolved by our setup that has a resolution of 400 ps, the fluorescence from the low energy peak can be measured as it has a much longer lifetime. The normalized temporal response from the low energy peak fits well with a double exponential function: $I(t) \sim a_M e^{\frac{-t}{\tau_M}} + a_{Def} e^{\frac{-t}{\tau_{Def}}}$. The fit returns $\tau_M = 680$ ps, $a_M = 0.75$, $\tau_{Def} = 3.4$ ns, and $a_{Def} = 0.13$ indicating the presence of two contributions to the low energy peak: a predominant contribution with a lifetime of 680 ps and a minor contribution with a longer lifetime of 3.4 ns. The latter is consistent with the recombination from localized excitons in WS$_2$,(*29*) and the small amplitude corroborates the hypothesis of a weak contribution of the defective band in the emission from the low energy peak.

Circular polarization is a key property of excitons in TMDs that enables the study of new many-body phenomena, and, at the same time, can be employed to differentiate exciton species, including free and localized *A* excitons.(*33*) Due to the role of the 1H conduction band edge at the K valley in the optical transition of the *M* exciton, we expect the *M* exciton to be circularly polarized similarly to the *A* exciton. This prediction is confirmed by polarization-resolved PL measurements at T=8K shown in Fig. 4e in which we measure the emission spectra of WS$_2$ excited with $\sigma^+$ circular polarization and collected with $\sigma^+$ (red line) and $\sigma^-$ (black line). We measure a degree of polarization, $\rho = \frac{I^+ - I^-}{I^+ + I^-}$ (where $I^+$ and $I^-$ are the emission intensity of $\sigma^+$ and $\sigma^-$ polarized light, respectively), for the *M* exciton of ~20%.

**Conclusions**

In summary, we present a new scalable approach to engineer additional optical resonances in the spectrum of TMD systems. By taking advantage of the stable intermediate steps of the 1H → 1T phase transition process, we have theoretically and experimentally shown that atomic-size mosaics of mixed-phase patches can be obtained from pristine 1H samples by triggering phase changes via irradiation with short bursts of plasma at mild power. The resulting atomic structures are robust and can be used in heterostructures and device assemblies. This atomic manipulation of the TMD lattice induces the formation of additional electronic bands that give rise to optical transitions in part of the spectrum that were previously out-of-reach. Moreover, the new transitions are characterized by useful optoelectronic and photonic properties, including a strong emission, long radiative lifetime and circular polarization. This observation opens new scenarios for light harvesting applications, active metamaterials, optoelectronics and quantum materials.

AUTHOR INFORMATION

**Corresponding Author**

*Address correspondence to ggrosso@gc.cuny.edu

**Author Contributions**

Conceptualization: JMW (supporting), SBC (lead), GG (lead)

Methodology: JMW (supporting), SBJ (lead), JF (supporting), GG (supporting)

Software: JMW (lead), SBC (supporting), EM (supporting), JF (lead), GG (supporting)

Validation: JMW (lead), SBC (supporting), JF (supporting), GG (supporting)

Formal Analysis: JMW (lead), SBC (supporting), JF (supporting), GG (lead)

Investigation: JMW (lead), SBC (lead), AA (supporting), JF (supporting), GG (supporting)

Resources: TT (lead), KW (lead), JF (lead)

Data Curation: JMW (supporting), SBC (supporting), EM (lead), JF (supporting), GG (supporting)

Writing – original draft: JMW (lead), SBC (lead), EM (supporting), JF (supporting), GG (lead)

Writing – review & editing: JMW (lead), SBC (supporting), EM (supporting), JF (supporting), GG (lead)

Visualization: JMW (lead), SBC (supporting), JF (supporting), GG (supporting)

Supervision: JF (supporting), GG (lead)

Project administration: JF (supporting), GG (lead)

Funding acquisition: JF (supporting), GG (lead)

## Notes

The authors declare no competing financial interest.

## Acknowledgements

G.G. acknowledges support from the National Science Foundation (NSF) (grant no. DMR-2044281), support from the physics department of the Graduate Center of CUNY and the Advanced Science Research Center through the start-up grant, and support from the Research Foundation through PSC-CUNY award 64510-00 52. K.W. and T.T. acknowledge support from the Elemental Strategy Initiative conducted by the MEXT, Japan (Grant Number JPMXP0112101001) and JSPS KAKENHI (Grant Numbers 19H05790, 20H00354 and 21H05233). J.F. acknowledges support from the NSF through grant NSF-2216838 and the NSF Phase II CREST Center IDEALS (grant no. EES-2112550). The Flatiron Institute is a division of the Simons Foundation.

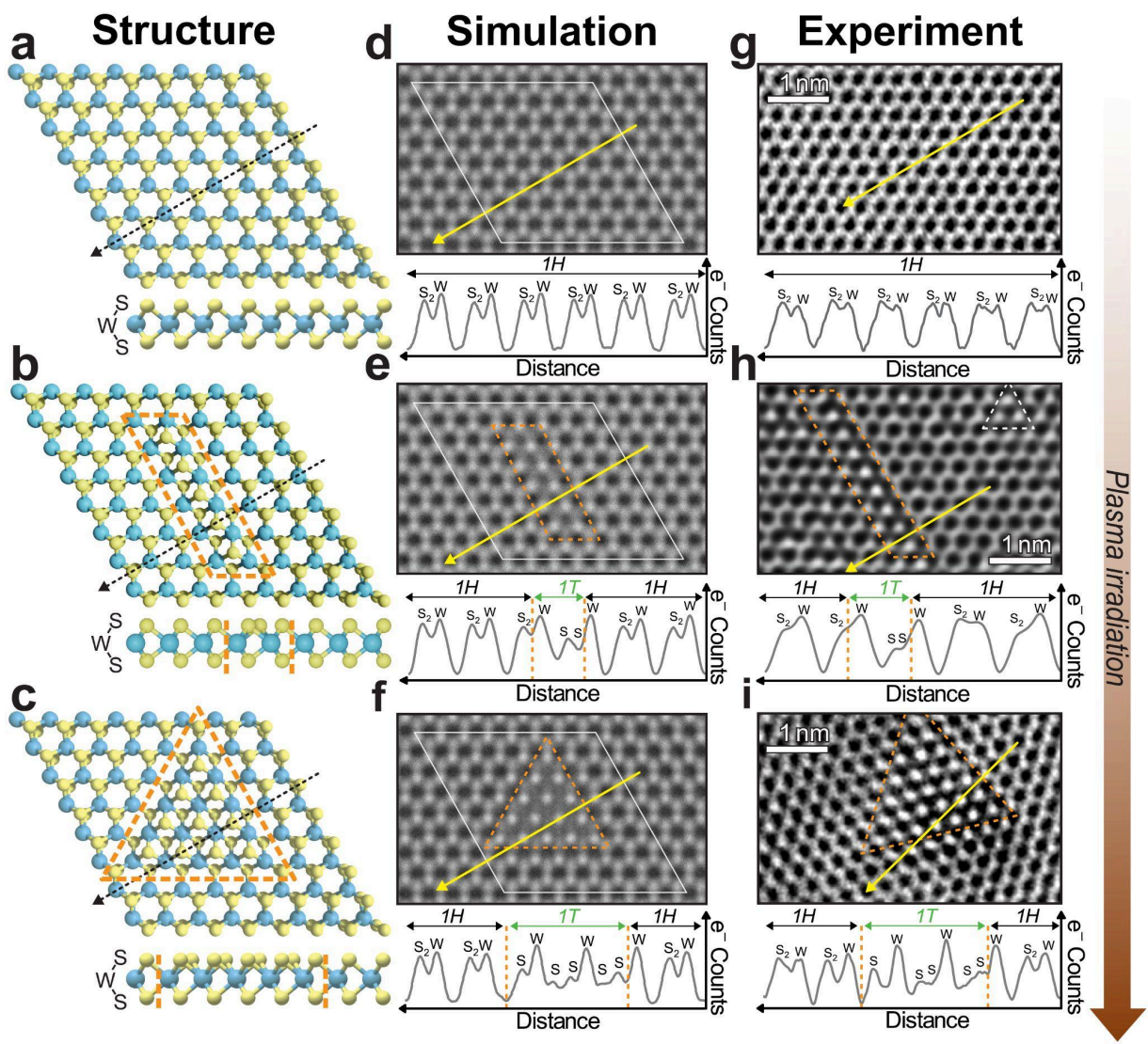

**Figure 1**. Top and side views of the atomic structure of (**a**) 1H pristine, (**b**) linear 1T/1H mixed phase structures (1T-L1), and (**c**) triangular 1T/1H mixed phase (1T-Tr4) WS$_2$ monolayers. Orange, dashed lines indicate the extent of 1T grain, and black, dashed lines indicate equivalent path to intensity line profiles in (d-i). Simulated TEM images (**d-f**) of the structures in (a-c), indicated by a thin white line, surrounded by extra pristine 1H-WS$_2$. Bottom trace of each panel indicates the intensity line profile along the path highlighted by the yellow arrow. There is a noticeable difference between S$_2$ stacks in 1H-WS$_2$ and isolated sulfurs in 1T-WS$_2$. Experimental TEM images of pristine WS$_2$ (**g**) and WS$_2$ flakes treated with argon plasma (see experimental section) for 5 s (**h**) and 10 s (**i**). The bottom trace of each panel is taken along the path indicated by the yellow arrow showing good agreement between the experiment and simulated TEM images. WS$_2$ treated for 5s primarily contains linear 1T grains (see Figure S3 for lower magnification TEM images), but we can also see small triangular grains indicated in (h) by a white dashed line. As the plasma treatment time is increased further, the phase mosaic is dominated by larger triangular and polygonal grains as demonstrated in (i).

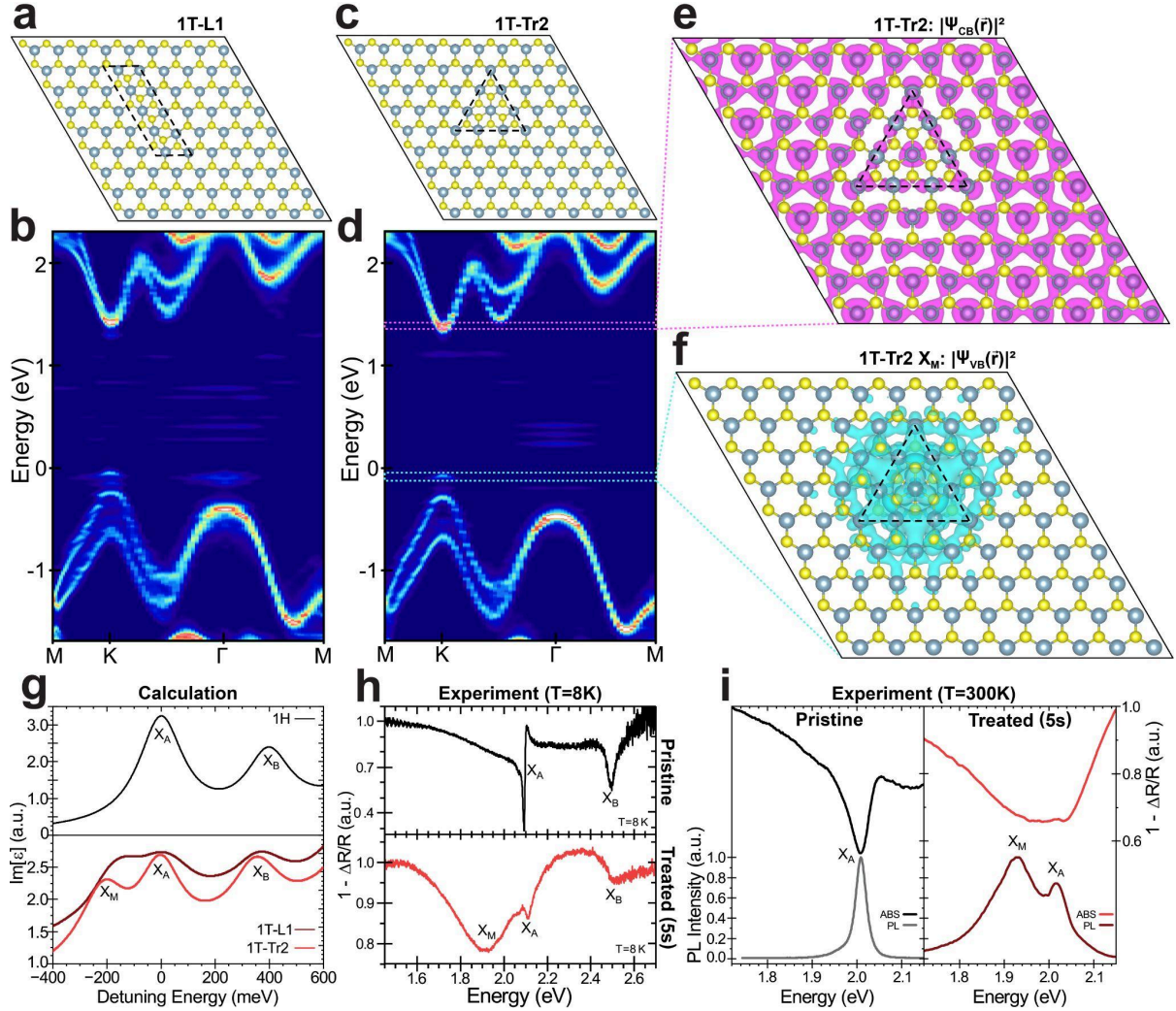

**Figure 2.** 9x9x1 relaxed supercell with linear 1T-L1 (**a**) and triangular 1T-Tr2 (**c**) structures of 1T phase within a $WS_2$ monolayer in the 1H phase. (**b, d**) Corresponding band structure calculated with DFT (see methods) for these mixed-phase supercells. The color scale in (**b, d**) represents the spectral weight of the band structure at a given momentum-energy coordinate. Realspace extent of the pseudo-wave function of the conduction (**e**) and valence (**f**) states involved in the *M* exciton transition for the 1T-Tr2 structure. The orbitals of the valence band are localized around the 1T phase, while the ones of the conduction band are delocalized across the 1H phase domain. (**g**) Calculated imaginary part of the dielectric function for 1H pristine (black curve), 1T-Tr2 (red curve), and 1T-L1 (maroon curve). The *M* exciton resonance ($X_M$) emerges below the *A* and *B* excitons. (**h**) Differential reflectivity measurements at T = 8K of pristine (black curve) and mixed-phase (red curve) samples treated with plasma for 5 seconds, indicating a good agreement with theoretical calculations of (**g**). (**i**) Comparison of the absorption (top plots) and emission (bottom plots) spectrum of pristine (black curves) and 5s-treated mixed-phase (red curves) samples at room temperature. Again, the *M* excitons appear below the *A* exciton in both absorption and emission spectra.

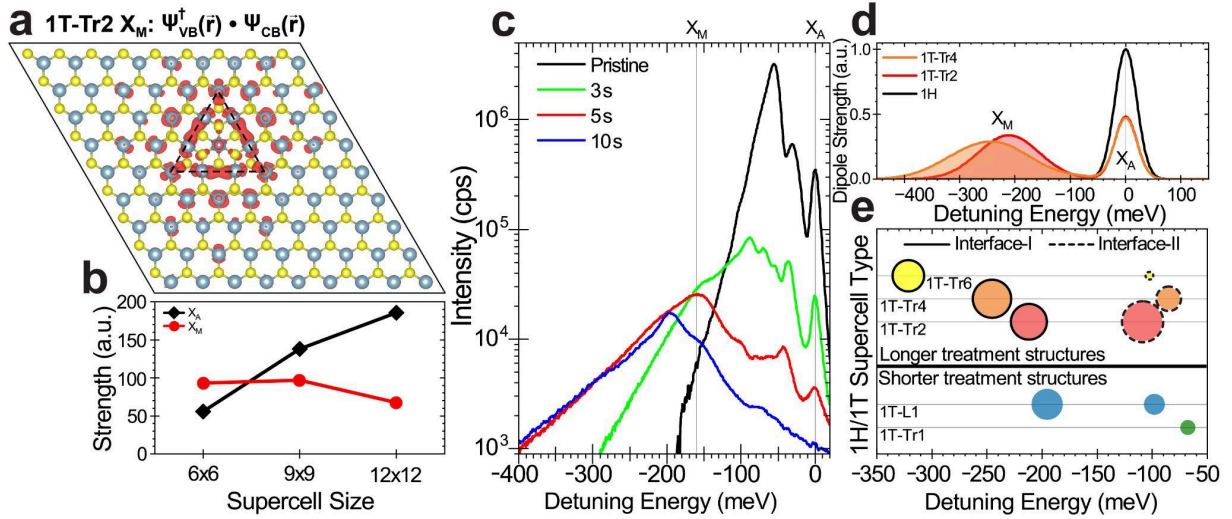

**Figure 3**. (**a**) Calculated transition density of the *M* exciton in 1T-Tr2 supercell which is concentrated along the interface between the triangular 1T grain and the surrounding 1H phase. (**b**) Calculated dipole strength of $X_A$ and $X_M$ resonances in 1T-Tr2 supercells of increasing size. The strength of $X_A$ grows as larger supercells have more 1H-$WS_2$. In contrast the strength of $X_M$ remains almost constant throughout, a further indication of the interfacial nature of the *M* exciton. (**c**) Emission spectra at T=8K taken with the same excitation laser power of pristine and mixed-phase samples treated for 3, 5, and 10 seconds. Spectra are indexed at the energy of the *A* exciton. (**d**) Plot of the calculated dipole strength as a function of detuning energy for $X_A$ and $X_M$ in 1H (black), 1T-Tr2 (red), and 1T-Tr4 (orange) supercells. For legibility, constituent transition dipoles are represented by a Gaussian distribution. (**e**) Summary of the energy and strength of the transition dipoles extracted from DFT calculations on 9x9x1 supercells containing different atomic-sized mixed-phase structures. Size of bubbles indicates the strength of the $X_M$ resonance, represented in **d** as the area under the $X_M$ peak. Dashed circles highlight the results of mixed-phase structures with type II interfaces. The dipole transition energy is plotted as a function of the detuning energy from the *A* exciton. An overall redshift of the *M* exciton occurs as a function of the size of the 1T grain, in agreement with the experimental data.

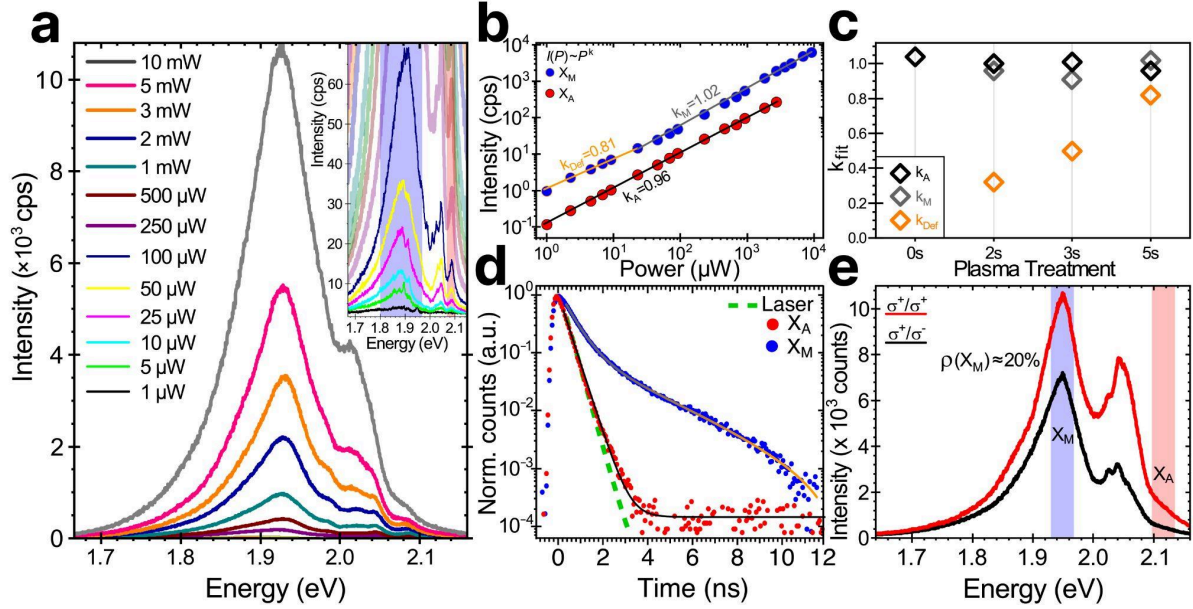

**Figure 4:** (**a**) – Emission spectra at increasing excitation laser power of a monolayer $WS_2$ treated for 5 s at 25 W. The *M* exciton peak at ~1.93 eV does not show any saturation behavior and dominates the spectrum at high power. Inset is a zoom at the spectral emission at low power. (**b**) PL intensity as a function of laser power for the *A* (red dots) and *M* (blue dots) excitons taken from the shaded windows of the inset in (**a**). Data are fitted with a power function $I(P)=aP^k$. At low power, the fit (orange line) of low-energy emission indicates defect-localized emission of the *A* exciton; however, at higher powers (gray line), this is eclipsed by excitonic emission. (**c**) Summary of the power coefficients for the defect-localized (orange), *M* excitons (gray), and *A* excitons (black) for increasing plasma treatment time. (**d**) Comparison of the fluorescence lifetime of *A* (red dots) and *M* (blue dots) excitons. The lifetime of the *A* excitons in the order of a few ps cannot be resolved with the time resolution (~400 ps) of our setup, and its temporal response returns a lifetime comparable to that of the laser (green dashed curve). The fluorescence lifetime of the *M* exciton fits well with a double exponential function with $\tau_M$=680 ps and $\tau_{Def}$=3.4 ns. (**e**) Emission spectra of another 5 s plasma treated $WS_2$ sample taken at T = 8 K excited with $\sigma^+$ circular polarization (with 500 μW laser excitation) and collected with $\sigma^+$ (red line) and $\sigma^-$ (black line).